\documentclass[lettersize,journal]{IEEEtran}
\usepackage{algorithm}
\usepackage{algpseudocode}
\usepackage{caption}
\usepackage{xcolor}
\usepackage[most]{tcolorbox}
\usepackage{booktabs}
\usepackage{amsmath,amsfonts}
\usepackage{array}
\usepackage[caption=false,font=normalsize,labelfont=sf,textfont=sf]{subfig}
\usepackage{textcomp}
\usepackage{stfloats}
\usepackage{url}
\usepackage{verbatim}
\usepackage{graphicx}
\usepackage{hyperref}    
\usepackage{cite}

\usepackage{multirow}
\usepackage[flushleft]{threeparttable} 

\usepackage[normalem]{ulem}
\useunder{\uline}{\ul}{}
\usepackage{xurl} 

\newcolumntype{P}[1]{>{\centering\arraybackslash}p{#1}} 
\newcolumntype{M}[1]{>{\centering\arraybackslash}m{#1}} 

\def\BibTeX{{\rm B\kern-.05em{\sc i\kern-.025em b}\kern-.08em
    T\kern-.1667em\lower.7ex\hbox{E}\kern-.125emX}}
\usepackage{balance}
\begin{document}




\title{CyberMentor: AI Powered Learning Tool Platform to Address Diverse Student Needs in Cybersecurity Education}



\author{Tianyu~Wang\href{https://orcid.org/0000-0002-9244-8798}{\includegraphics[scale=0.6]{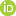}}, Nianjun~Zhou\href{https://orcid.org/0000-0002-3473-6097}{\includegraphics[scale=0.6]{images/orcid_16x16.png}}, and Zhixiong~Chen\href{https://orcid.org/0000-0002-9874-6972}{\includegraphics[scale=0.6]{images/orcid_16x16.png}}
\thanks{Tianyu Wang is with Mercy University, Math \& Computer Science Department, 555 Broadway, Dobbs Ferry, NY 10522, USA (e-mail: twang4@mercy.edu).}
\thanks{Nianjun Zhou is with  IBM Research AI
of T.J. Watson Research Center, 1101 Kitchawan Rd, Yorktown Heights, NY 10598, USA (e-mail: jzhou@us.ibm.com).}
\thanks{Zhixiong Chen is with Mercy University, Math \& Computer Science Department, 555 Broadway, Dobbs Ferry, NY 10522, USA (e-mail: zchen@mercy.edu).}}

\markboth{Journal of \LaTeX\ Class Files,~Vol.~xx, No.~xx, xxxx}%
{How to Use the IEEEtran \LaTeX \ Templates}

\maketitle
\begin{abstract}

Many non-traditional students in cybersecurity programs often lack access to advice from peers, family members and professors, which can hinder their educational experiences. Additionally, these students may not fully benefit from various LLM-powered AI assistants due to issues like content relevance, locality of advice, minimum expertise, and timing. This paper addresses these challenges by introducing an application designed to provide comprehensive support by answering questions related to knowledge, skills, and career preparation advice tailored to the needs of these students. We developed a learning tool platform, \textit{CyberMentor}\footnote{https://github.com/tisage/CyberMentor}, to address the diverse needs and pain points of students majoring in cybersecurity. Powered by agentic workflow and Generative Large Language Models (LLMs), the platform leverages Retrieval-Augmented Generation (RAG) for accurate and contextually relevant information retrieval to achieve accessibility and personalization. We demonstrated its value in addressing knowledge requirements for cybersecurity education and for career marketability, in tackling skill requirements for analytical and programming assignments, and in delivering real time on demand learning support. Using three use scenarios, we showcased CyberMentor in facilitating knowledge acquisition and career preparation and providing seamless skill-based guidance and support. We also employed the LangChain prompt-based evaluation methodology to evaluate the platform’s impact, confirming its strong performance in helpfulness, correctness, and completeness. These results underscore the system’s ability to support students in developing practical cybersecurity skills while improving equity and sustainability within higher education. Furthermore, CyberMentor’s open-source design allows for adaptation across other disciplines, fostering educational innovation and broadening its potential impact.

\end{abstract}


\thanks{Copyright \copyright\ 2024 IEEE. Personal use of this material is permitted. Permission from IEEE must be obtained for all other uses, in any current or future media, including reprinting/republishing this material for advertising or promotional purposes, creating new collective works, for resale or redistribution to servers or lists, or reuse of any copyrighted component of this work in other works.}

\section{Introduction}\label{sec:introduction}


In higher education, the scarcity of educational resources always poses significant challenges throughout a student's academic journey, particularly for non-traditional and minority students who bear significant financial or familial responsibilities \cite{horn1996nontraditional}. These students often face increased difficulties in accessing critical advice from peers, professors, and other learning and research resources, which can exacerbate the challenges they encounter in their educational experiences \cite{jones1990high}. To address these specific challenges, a comprehensive solution that provides tailored support, encompassing key areas such as knowledge acquisition, skill development, career preparation, and on-demand support, becomes more critical for them.

\begin{figure}[ht]
\centering
\includegraphics[width=8.5cm]{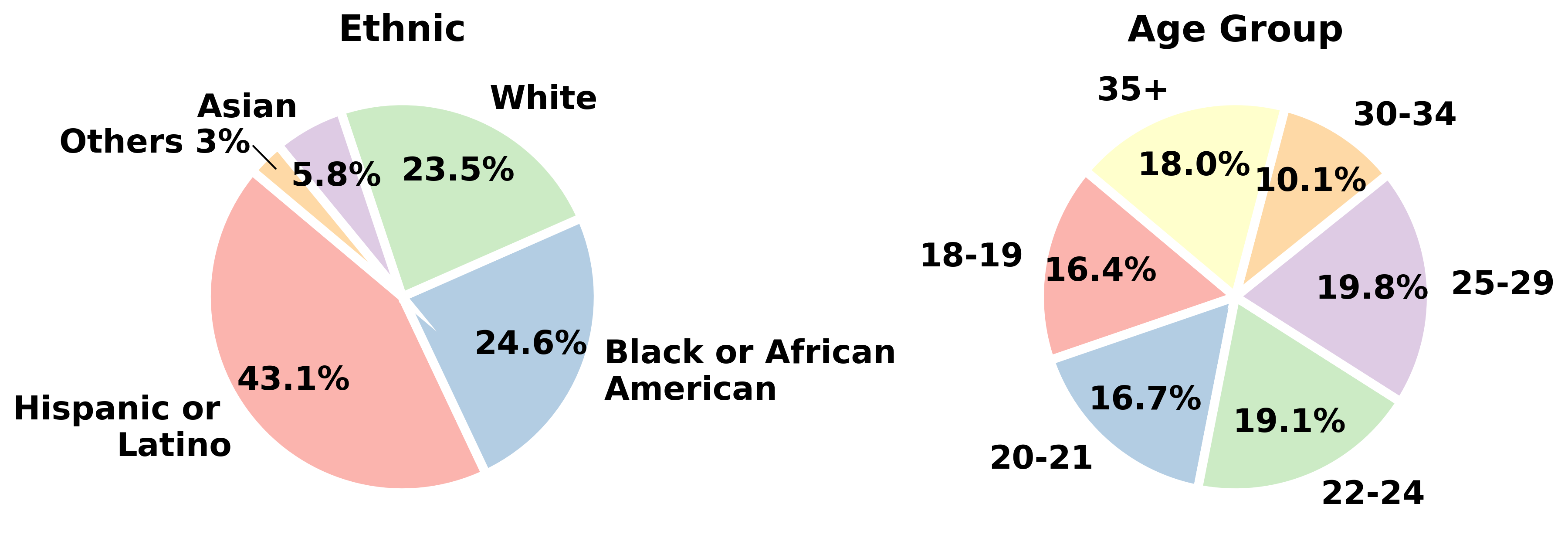}
\caption[Caption for LOF]{Demographic Distribution (Ethnic \& Age Groups)\protect\footnotemark.}
\label{fig:DemographicDistribution}
\end{figure}

\footnotetext{https://www.collegefactual.com/colleges/mercy-college/student-life/diversity}

\begin{figure}[ht]
\centering
\includegraphics[width=8.5cm]{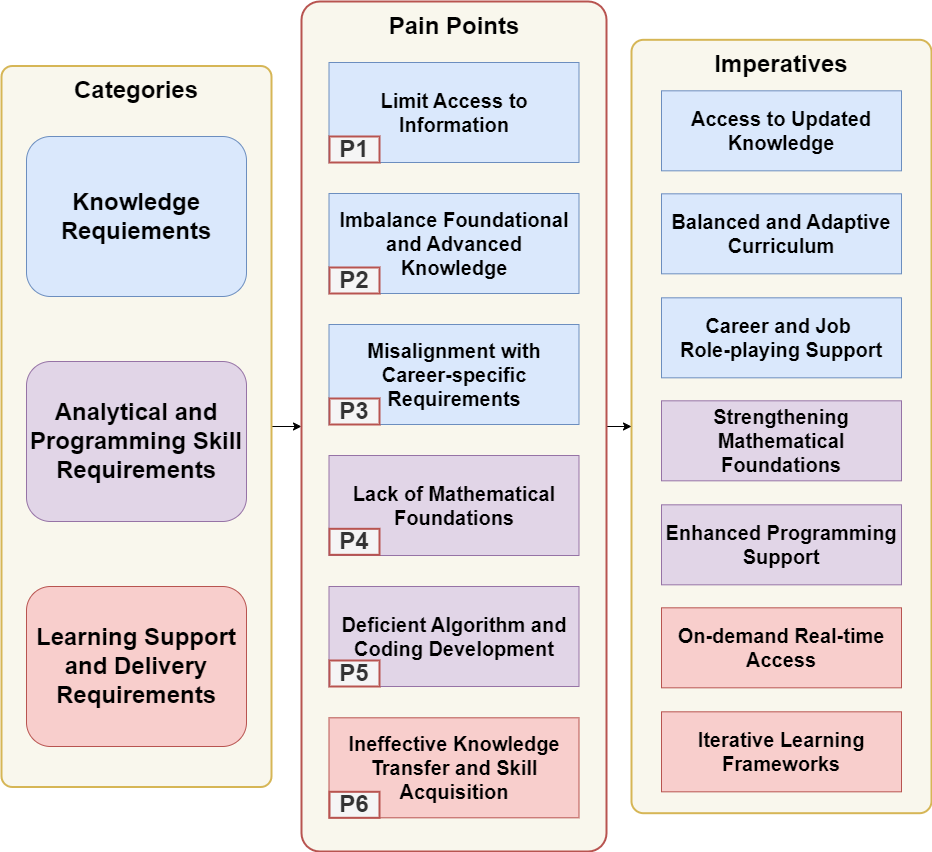}
\caption{Six Identified Pain Points in Cybersecurity Education}
\label{fig:PainPoint}
\end{figure}

Our study began with an in-depth examination of the pain points experienced by students majoring in cybersecurity, followed by an analysis of the student demographics that reveal such challenges in education-focused universities. Cybersecurity was selected as the focus due to its unique and evolving demands. The field requires the mastery of a broad and rapidly expanding knowledge base, coupled with strong analytical skills. Its dynamic nature—shaped by emerging threats, frequent security policy updates, and evolving career opportunities—necessitates continuous learning and adaptability beyond the current curriculum. These characteristics make cybersecurity an ideal domain for exploring the feasibility and impact of generative AI technologies in educational contexts. At our university, which is typical of many educational institutions, a significant portion of the student population comprises commuters, part-time students, and veterans. This demographic composition reflects a broader trend in higher education, where non-traditional students are becoming increasingly common\footnote{https://www.luminafoundation.org/campaign/todays-student/}. Figure~\ref{fig:DemographicDistribution} illustrates the demographic distribution, highlighting the diversity within the student body. For a university characterized by a diverse student body and a high percentage of part-time enrollees, a persistent challenge—common not only in cybersecurity but also in STEM education — lies in bridging the gap between educational demands and available resources. Students in this discipline, coming from varied backgrounds, require flexible educational solutions that include adaptive mentoring, adjustable teaching schedules, and accessible support beyond traditional classroom hours. Figure~\ref{fig:PainPoint} categorizes the primary challenges faced by these students into three main areas: 
\begin{enumerate}
    \item \textbf{Mentorship and Accessibility Issues}: Limited access to specialized mentors and rigid course schedules hinder personalized learning opportunities.
    \item \textbf{Outdated Knowledge Resources}: A rapidly evolving field often leaves students struggling to keep pace with up-to-date information.
    \item \textbf{Career Support and Certification Gaps}: Insufficient time and resources for certification preparation and career readiness impede students’ ability to achieve their professional goals.
\end{enumerate}
These challenges are further exacerbated by the limited availability of specialized mentors and the constraints of rigid schedules, which together hinder students' ability to align their education with their aspirations. Our study identifies these barriers as critical areas for intervention, providing the foundation for our proposed solution.

To address these challenges, we propose \textit{CyberMentor}, a virtual mentoring framework. \textit{CyberMentor} integrates multiple virtual instructors through a centralized access point, facilitating personalized learning experiences tailored to individual needs. This framework is designed to accommodate learners from diverse educational backgrounds and life circumstances, supporting their academic pursuits, research endeavors, and career development queries. The key contributions of this work are as follows:

\begin{enumerate}
    \item \textbf{Identification and Analysis of Pain Points in Cybersecurity Education:} 
    We identify and analyze six major challenges faced by students and provide essential recommendations to enhance accessibility and engagement in cybersecurity education.

    \item \textbf{Agentic Workflow for Natural Language Interaction:} 
    This framework incorporates an agentic workflow to facilitate natural language interactions, making the system more intuitive and accessible for students. 

    \begin{itemize}
    \item \textbf{Cybersecurity Knowledge Integration:} 
    It integrates comprehensive domain-specific knowledge using Retrieval-Augmented Generation (RAG), covering areas such as threats and vulnerabilities, security practices, risk management, and compliance with frameworks and standards like NIST and GDPR. It also includes emerging technologies like Artificial Intelligence (AI), Internet-of-Things (IoT) security, and the legal and ethical aspects of cybersecurity, along with career development in these fields.

    \item \textbf{Tools for Advanced Skill Development:} 
     It supports the development of advanced technical skills that are critical for professional success in the cybersecurity domain. Through a series of integrated tools and resources, students can gain expertise in areas such as cryptography, anomaly detection, malware analysis, and programming for secure applications.

    \end{itemize}

\end{enumerate}

In the following sections, we will explore the challenges faced by students in cybersecurity education and the necessary imperatives to address them. Subsequently, a literature review will cover existing research on cybersecurity education and the integration of LLMs in higher education. We will then introduce our proposed framework, which addresses these challenges through various approaches, including the implementation of RAG and the iterative learning process for analytical skills within the agentic workflow, as well as the application of coding and machine learning algorithms for cybersecurity detection. Finally, we will validate our framework, discuss the potential and limitations of Generative AI and LLMs in education, and conclude our study.

\section{Challenges in Cybersecurity Education with a Diverse Student Body}

In a cybersecurity education program, particularly at institutions with a diverse student body, it is crucial to address the varied educational needs of students to ensure equitable learning outcomes. These needs span acquiring foundational and advanced cybersecurity knowledge, developing analytical and programming skills, and overcoming the unique challenges students face due to their diverse backgrounds. This section delves into the specific requirements and associated pain points students encounter in their academic journey. We explore how these challenges can be addressed through targeted educational imperatives, as Fig. \ref{fig:PainPoint} shown, setting the stage for the subsequent discussion on implementing a scalable solution to meet these needs effectively.

\subsection{Knowledge Requirements for Cybersecurity Education}
This subsection explores the comprehensive knowledge base required for students pursuing cybersecurity education. To develop a strong foundation, students must gain proficiency in key areas of the field, such as network security, cryptography, threat intelligence, and risk assessment. Mastery of these foundational areas is crucial for students to understand the broad spectrum of cybersecurity roles\footnote{https://public.cyber.mil/wid/pathways}. For instance, roles such as Information Systems Security Developer and Cyber Defense Analyst  are pivotal for tasks like cryptography development, detecting network anomalies, and identifying misinformation. These roles demand not only deep knowledge within their respective domains but also continuous learning and adaptation to emerging technologies, threats, and industry standards. Consequently, students encounter several challenges that may impede their ability to fully engage with and master these critical concepts:

\begin{itemize}
    \item \textbf{\textit{P1: }Limited Access to Information:} A primary challenge is difficulty accessing the most current and relevant information. The fast-paced changes in the cybersecurity landscape necessitate that students and educators continuously adapt to new threats, policies, and tools. However, updating educational materials is resource-intensive, requiring significant instructor effort to gather and properly prepare the latest information for student consumption.

    \item \textbf{\textit{P2: }Imbalance Balancing Foundational and Advanced Knowledge:} Another challenge lies in balancing providing a solid foundational understanding and advancing specialized knowledge. Students from diverse backgrounds may struggle to keep pace with both fundamental concepts and the latest developments in cybersecurity. Instructors also face the challenge of balancing course difficulty to ensure that students grasp the essential knowledge while maintaining their enthusiasm for learning.

    \item \textbf{\textit{P3: }Misalignment with Career-Specific Requirements:} Students must tailor their learning to align with their intended career paths, which can be difficult without clear guidance and access to targeted resources. The need to adjust learning strategies based on specific career goals adds further complexity to their educational experience.
\end{itemize}

To effectively mitigate these challenges in cybersecurity education, the following imperatives are essential:

\begin{itemize}
    \item \textbf{Access to Updated Knowledge:} Educational services must provide timely updates to ensure that students are learning the latest industry standards and practices.

    \item \textbf{Balanced and Adaptive Curriculum:} Educational programs must strike a careful balance between providing a strong foundational understanding and offering exposure to advanced, specialized knowledge. To achieve this, the curriculum should integrate academic resources, industry publications, and practical experiences, while remaining adaptable to students' varying levels of prior knowledge to ensure personalized learning pathways that keep pace with the evolving field.

    \item \textbf{Career and Job Role-playing Support:} Educational services must also offer clear guidance and resources tailored to the specific career paths students wish to pursue. This includes providing targeted modules, aligning learning outcomes with industry needs, workshops, and mentoring to help students focus on the skills and knowledge most relevant to their intended roles in the industry.
\end{itemize}

\subsection{Analytical and Programming Skill Requirements}
Another significant challenges encountered by students in cybersecurity education is the requirement to develop a robust foundation in both analytical and programming skills. These challenges can be broadly categorized into several key areas:

\begin{itemize}
    \item \textbf{\textit{P4: }Lack of Mathematical Foundations:} A solid foundation in mathematics is critical for understanding and applying concepts in cybersecurity, particularly in cryptography and data analysis. However, many students struggle with advanced mathematical topics due to gaps in their prior knowledge. This lack of a strong mathematical foundation can severely limit their ability to grasp complex concepts, thereby hindering their overall academic progress.

    \item \textbf{\textit{P5: }Deficient Algorithm  and Coding Development:} The development of effective algorithms and their implementation through coding necessitate strong problem-solving abilities, a deep understanding of algorithmic thinking, and proficiency in programming languages. Students often find these challenges daunting due to the abstract nature of algorithmic reasoning and the iterative process of coding and debugging required to refine solutions. The steep learning curve can be particularly overwhelming for students from non-technical backgrounds, leading to frustration and hindering their progress in acquiring essential cybersecurity skills. 
\end{itemize}

To address these skill gaps, educational services must implement targeted strategies to support student success, guided by the following imperatives:

\begin{itemize}
    \item \textbf{Strengthening Mathematical Foundations:} Educational services must provide comprehensive resources tailored to individual needs. This should include personalized tutoring sessions, a wide range of practice exercises, and targeted instructional support aimed at reinforcing mathematical concepts essential to cybersecurity. By strengthening these foundations, students can better grasp advanced topics and apply mathematical reasoning to solve complex cybersecurity challenges.

    \item \textbf{Programming Support with Iterative Frameworks:} Educational services should implement comprehensive support mechanisms by combining personalized tutoring with an iterative learning framework. Personalized tutoring must adapt to individual learning paces, offering tailored guidance and feedback, while iterative learning promotes continuous practice and refinement in coding and problem-solving, helping students gradually develop algorithmic thinking and the resilience needed to confidently tackle complex cybersecurity tasks.

\end{itemize}

\subsection{Learning Support and Delivery Requirements}
The effectiveness of cybersecurity education relies not only on the curriculum content but also on the delivery and accessibility of educational services. A major challenge can be identified:

\begin{itemize}
    \item \textbf{\textit{P6: } Ineffective Knowledge Transfer and Skill Acquisition}. A significant challenge in cybersecurity education, especially in non-research institutions, is the limited availability of mentoring resources. Students often find themselves struggling to access timely and personalized guidance, which is crucial for mastering complex cybersecurity concepts and skills. This scarcity of mentoring is particularly acute when it comes to developing hands-on abilities, such as coding, configuring security tools, or conducting forensic analyses. The absence of sufficient mentoring and support can lead to gaps in knowledge, decreased motivation, and a lack of confidence in applying theoretical knowledge to practical scenarios.
\end{itemize}

To effectively enhance educational service delivery, the following imperatives strategies should be implemented:

\begin{itemize}
    \item \textbf{On-demand Real-time Access:} To mitigate the shortage of instructor availability, it is essential to implement a low-cost, scalable, 24/7 real-time educational support system that provides continuous assistance. Such a system would enable students to promptly resolve their queries at any time, enhancing their learning experience, maintaining engagement, and reducing the stress of studying complex cybersecurity topics in the absence of human instructors.
    \item \textbf{Interactive Learning Framework:} Beyond real-time access, it is crucial to implement an interactive, iterative mentoring system that enables ongoing, multi-round interactions where students receive continuous feedback on cybersecurity concepts. By allowing students to ask questions in a natural, conversational manner and receive iterative responses, this system would support personalized learning, helping them assess their progress and refine their study strategies for a deeper and more effective learning experience.
\end{itemize}



\section{Literature Review}


Recent literature highlights a significant shift in the demographic composition of students in higher education, with a growing percentage of non-traditional students\footnote{https://nces.ed.gov/programs/coe/indicator/ctu
}. These students, who often struggling their studies with other responsibilities such as work or family, represent a diverse population that includes part-time learners, older students, veterans, and those pursuing education later in life \cite{center2015yesterday}. This demographic shift has important implications for educational institutions, particularly in fields like cybersecurity, where the demands of the curriculum can be intense. Non-traditional students often face unique challenges, such as limited time for study, less access to peer support, and a need for more flexible learning options\cite{altbach2011american}. The literature suggests that educational programs must adapt to these changing demographics by offering more personalized and flexible learning environments \cite{ross2011research, wladis2015stem, beckwith2023barriers}.


Cybersecurity education faces several significant challenges, particularly in programs with catering to career-changing professionals and diverse student populations. The rapid evolution of technology and threat landscapes makes it difficult for students to keep pace with the latest industry developments, resulting in a disconnect between academic instruction and real-world requirements \cite{singer2014cybersecurity, santos2017challenges, shillair2022cybersecurity, triplett2023addressing}. Additionally, the inherent complexity of cybersecurity concepts, such as cryptography and network security, imposes a high cognitive load, particularly for students from non-technical backgrounds \cite{santos2017challenges, triplett2023addressing}. Another critical issue is the integration of hands-on experience with theoretical knowledge, as practical skills are indispensable for addressing real-world cybersecurity problems. Unfortunately, access to hands-on labs, simulations, and real-world scenarios remains limited, especially for non-traditional students, exacerbating the challenge of achieving a balance between theory and application in cybersecurity education \cite{gasiba2020design, john2020cybersecurity}. These challenges underscore the need for innovative educational solutions to bridge the gap between academic learning and industry demands.

In response to the challenges facing cybersecurity education, current pedagogical practices have evolved to incorporate a range of approaches aimed at bridging the gap between theoretical knowledge and practical application. Many programs now emphasize integrated learning models that combine case studies, simulations, and capstone projects, allowing students to apply theoretical concepts in realistic, hands-on environments \cite{aldaajeh2022role}. The rise of online and hybrid education models has further expanded access, particularly for non-traditional students, by offering flexible, asynchronous learning options that accommodate diverse schedules and learning paces \cite{paar2010understanding, rahman2020importance, ros2020analyzing}. Additionally, collaboration with industry partners has become increasingly common, with institutions working closely with companies to align curricula with evolving industry demands. This includes providing internships, co-op programs, and industry-led workshops that enhance students' readiness for the workforce \cite{aldaajeh2022role, blavzivc2022changing}.

Various enabling technologies have also been explored. Virtual labs and simulations provide students with hands-on experience in controlled environments, offering significant benefits to non-traditional students who may lack access to physical lab facilities \cite{robles2020emulating, workman2021study, vykopal2021scalable}. Advanced learning management systems (LMS) are also widely employed to manage course content, track student progress, and facilitate communication between students and instructors, playing a critical role in the successful implementation of online and hybrid education models \cite{alexei2021cyber, krumova2023education}. Additionally, emerging technologies such as artificial intelligence (AI) and machine learning (ML) are being utilized to create adaptive learning systems capable of tailoring educational content to individual student needs. These systems provide personalized feedback, recommend resources, and adjust task difficulty based on student performance \cite{laato2020ai, kallonas2024empowering}. Moreover, technologies like Retrieval-Augmented Generation (RAG) and large language models (LLMs) are being leveraged to deliver real-time, up-to-date information, helping students stay current with evolving cybersecurity trends \cite{xu2024large, lewis2020retrieval, li2024chainthoughtempowerstransformers}. These tools further support automated grading, personalized content delivery, and the provision of instant feedback, enhancing the overall learning experience \cite{10386406, simoni2024morse, shen2024hugginggpt, wang2024tools, snell2024scaling, block2014mystery}.

\section{Solution Outline}\label{sec:solution_outline}

In response to the evolving challenges in cybersecurity education, we propose a scalable LLM-based mentoring framework\footnote{https://github.com/tisage/CyberMentor}, termed \textit{CyberMentor}. This innovative chatbot system is designed to bridge the gap between student and instructor's needs and available educational resources, offering a low-cost, 24/7, multilingual, personalized and effective learning support. By leveraging open-source LLM frameworks, \textit{CyberMentor} provides tailored assistance for student queries, encompassing career guidance, knowledge units, analytical questions, and technical support across various cybersecurity field.

\subsection{Solution Components}

\begin{figure*}[ht]
\centering
\includegraphics[width=17cm]{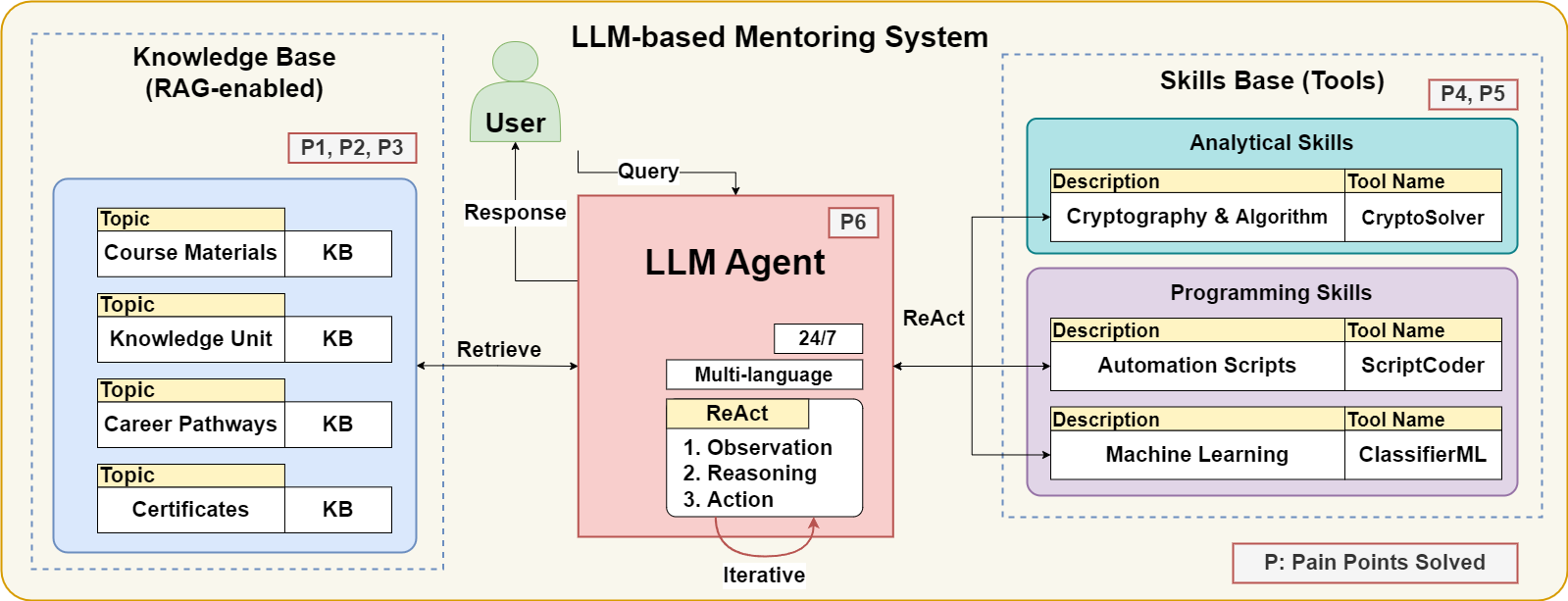}
\caption{Proposed Framework (\textit{CyberMentor}) to Enhance Cybersecurity Education. (P1-P6 refer to Pain Points in Fig. \ref{fig:PainPoint})}
\label{fig:Model}
\end{figure*}

As shown in Fig \ref{fig:Model}, our model consists of three major components:

\begin{enumerate}
    \item \textbf{Knowledge Base (KB)}: The Knowledge Base is an organized repository that encompasses a wide array of educational resources within the cybersecurity domain, including course learning materials, knowledge units, career pathways, and certification materials. This component primarily addresses pain points related to accessing updated information, balancing foundational and advanced knowledge, and career-specific knowledge adjustment, as depicted in Fig.\ref{fig:PainPoint} (P1, P2, P3). A detailed methodology and implementation strategy for the Knowledge Base are discussed in Section\ref{section:knowledge_base}.

    \item \textbf{Skill Base (Tools)}: The Skill Base is a curated suite of technical tools designed to address specific categories of challenges within the cybersecurity field. These tools not only provide hands-on problem-solving capabilities but also incorporate interactive learning modules that are crucial for mastering practical cybersecurity skills. Additionally, the tools are fully customizable, allowing educators and students to tailor the learning experience to meet specific educational goals and individual needs. Addressing pain points such as the need for strong mathematical foundations, algorithm development, and programming challenges, the Skill Base caters to both the analytical and programming skill development of students, as indicated in Fig.\ref{fig:PainPoint} (P4, P5). Further details on the functionalities and technical implementation of the Skill Base tools are provided in Section\ref{section:math} and Section\ref{section:coding}.

    \item \textbf{LLM Agent}: As proposed by Yao et al. \cite{yao2023reactsynergizingreasoningacting}, this agentic framework allows LLMs to generate reasoning traces that are interleaved with task-specific actions, effectively synergizing these processes to enhance decision-making and problem-solving capabilities. The LLM Agent functions as the central AI component of our proposed framework, tasked with delivering precise and contextually relevant support to students and solve the pain point related to the efficient and personalized learning experience, as illustrated in Fig. \ref{fig:PainPoint} (P6). By intelligently processing user queries, the LLM Agent selects the most suitable resources or tools from the Knowledge Base and Skill Base to provide tailored responses. This approach ensures that students receive accurate assistance using an interactive chat system that is aligned with their specific learning needs. It also employs 24/7 and multi-language processing capabilities, enabling it to interact with students in their native languages, thereby reducing language-related learning barriers and fostering an inclusive educational environment. This is particularly beneficial for non-traditional students and first-generation learners, who often face additional challenges in their academic journey. 

\end{enumerate}

More detailed real-world examples of these use cases are discussed in Section \ref{sec:use_scenarios}. The subsequent section will elaborate on the preparation of the knowledge base and skill-based tools that underpin our framework.

\section{Knowledge Base Implementation}\label{section:knowledge_base}
In fast-evolving fields like cybersecurity, where new threats and techniques emerge regularly, searching for relevant information that reflects the latest knowledge is crucial. Retrieval-Augmented Generation (RAG) is a hybrid framework that integrates retrieval mechanisms with generative models to enhance the generation of responses by incorporating relevant external information. While Large Language Models (LLMs) are trained on vast datasets, their knowledge is inherently limited to the data available at the time of training. As new cyber threats and techniques emerge continuously, relying solely on static training data may lead to outdated or incomplete information. Consequently, it is essential to retrieve up-to-date, domain-specific knowledge beyond the LLM's initial training data to ensure the model's outputs remain both relevant and accurate.

Unlike traditional databases that require predefined schema and structured queries, RAG dynamically retrieves and integrates information into the generative process. This method allows for more flexible, context-aware, and precise responses, making it particularly advantageous in educational settings where the context and needs of the learners can vary significantly. RAG’s ability to seamlessly combine retrieval with generation ensures that the information provided is both relevant and current, addressing the specific demands of cybersecurity education.

\begin{figure}[ht]
\centering
\includegraphics[width=8.5cm]{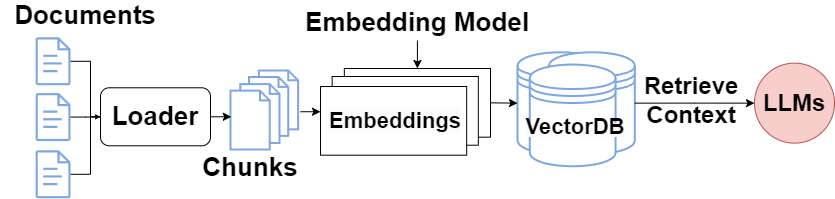}
\caption{Flowchart of Retrieval-Augmented Generation (RAG)}
\label{fig:RAG}
\end{figure}

\subsection{RAG Workflow}
The RAG workflow (Fig. \ref{fig:RAG}) comprises several critical steps to ensure the relevance and completeness of generated content. First, documents pertinent to the cybersecurity curriculum—such as knowledge units, career pathways, and program catalogs—are loaded into the system to provide comprehensive source information aligned with educational goals. These documents are then chunked into smaller, overlapping segments to maintain context between chunks, enabling precise retrieval without sacrificing coherence. Each chunk is converted into a vector representation using an embedding model, capturing the semantic meaning of the content for nuanced retrieval. The embeddings are stored in a vector database (\textit{VectorDB}), facilitating scalable and efficient retrieval of relevant chunks based on user queries or prompts to the LLM. This setup ensures fast and accurate retrieval, critical for providing real-time feedback as the volume of educational content grows. When students or educators interact with the system, the most relevant chunks are retrieved from the \textit{VectorDB} and fed into the LLM, ensuring responses are informed by up-to-date, context-specific information. Additionally, the LLM's responses may include references to original sources, such as URLs or specific sections in books, enhancing credibility and guiding users for further exploration.

\subsection{Documents Selection}

\begin{table}[ht]

\centering
\caption{Summary of Knowledge Base Categories and Their Corresponding Sources.}
\label{tab:knowledge_base}
\begin{threeparttable}
\begin{tabular}{c c}
\hline
\textbf{Knowledge Base} & \textbf{Source}                                                                                 \\ \hline
\multirow{2}{*}{Knowledge Units} & National Centers of Academic Excellence (NCAE-C)\tnote{a} \\ \cline{2-2} 
                        & CompTIA Security+ Certification Guide\cite{stewart2021comptia} \\ \hline
Career Pathways         & Federal Cyber Career Pathways Working Group\tnote{b}                           \\ \hline
Program Catalogs        & Cybersecurity Program Catalogs\tnote{c}                                        \\ \hline
\end{tabular}
\begin{tablenotes}
    \item[a] https://public.cyber.mil/ncae-c/documents-library/
    \item[b] https://public.cyber.mil/wid/pathways/
    \item[c] https://www.mercy.edu/academics/programs/bs-cybersecurity
\end{tablenotes}
\end{threeparttable}
\end{table}

In the context of RAG, documents are external data sources retrieved to enhance the output of LLMs. In this study, we have categorized documents into three primary types: Knowledge Units, Career Pathways, and Program Catalogs. For detailed information on the sources, and specific documents utilized, refer to Table \ref{tab:knowledge_base}. 

\begin{enumerate}
    \item \textbf{Knowledge Units}: These are essential for ensuring that students acquire the foundational skills and knowledge required in cybersecurity. We utilize documents from the National Centers of Academic Excellence (NCAE) and The Center of Academic Excellence in Cyber Defense (CAE-CD) and the CompTIA Security+ Certification Guide. These sources provide comprehensive frameworks and guidelines that support curriculum standards and competency development across various academic institutions.

    \item \textbf{Career Pathways}: These documents provide structured frameworks for career planning and skill development, crucial for improving student engagement and employment outcomes. Our research leverages materials from the Federal Cyber Career Pathways Working Group, aligned with the NICE Framework, to offer students clear guidance on potential career trajectories within the cybersecurity field.

    \item \textbf{Program Catalogs}: These catalogs offer detailed insights into academic programs, including curriculum, policies, and student services, specific to cybersecurity education. We incorporate catalogs from a selected university program to address the diverse needs of our student sample, ensuring relevance and applicability.
\end{enumerate}

\section{Analytical Skills Toolkit Using Agentic Workflow}\label{section:math}

Cryptography, a cornerstone of cybersecurity, presents a unique set of challenges for students, particularly in the development of robust analytical and problem-solving skills. As highlighted in the previous section, students often struggle with acquiring and applying complex mathematical foundations and related algorithm. These pain points are directly related to the study of cryptography, where the need to understand and implement cryptographic algorithms demands a high level of mathematical proficiency and algorithmic thinking.

\begin{figure}[ht]
\centering
\includegraphics[width=8.5cm]{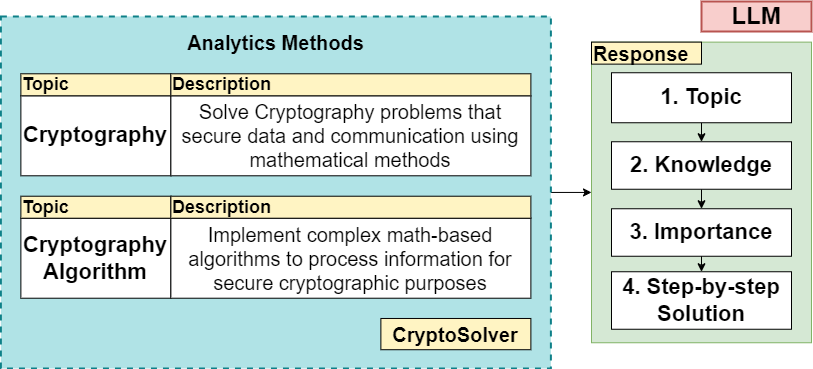}
\caption{A Skill-based Tool to Solve Cryptography Problem}
\label{fig:Analytics}
\end{figure}

Our proposed solution, the \textit{CryptoSolver} tool, part of \textit{CyberMentor} (Fig. \ref{fig:Analytics}), is an agentic workflow leveraging LLMs to address specific challenges in mastering cryptography. This workflow aligns with educational imperatives by facilitating a structured learning pathway that supports students through key steps: it begins with topic identification to pinpoint the specific cryptographic concept the student needs to focus on, ensuring targeted and relevant learning. It then provides knowledge acquisition by accessing and presenting the most up-to-date information on the topic, addressing the challenge of obtaining current cybersecurity knowledge. The workflow emphasizes importance clarification, highlighting the significance of the topic within the broader context to help students prioritize their learning efforts and adjust to career-specific requirements. Finally, it offers step-by-step solution development, providing detailed solutions to cryptographic problems to build students' confidence and competence in developing algorithms and applying mathematical methods. In summary, each step in the \textit{CryptoSolver} workflow is meticulously crafted to address the specific challenges previously identified. By offering a structured and adaptive learning pathway, this tool ensures that students can develop essential skills in a logical and cohesive manner.
\section{Coding Skills Toolkit Using Agentic Workflow}
\label{section:coding}

In the domain of cybersecurity, coding skills are also indispensable, particularly in automating tasks and leveraging machine learning techniques for advanced threat detection. Two pivotal aspects of coding in this domain are (1) the development of automated scripts for network anomaly detection and (2) the creation of machine learning models for sophisticated threat analysis. This section discusses how our agentic workflow tools, \textit{ScriptCoder} and \textit{MLClassifier}, address these essential needs by guiding students through structured processes that enhance their coding proficiency in cybersecurity contexts.

\begin{figure}[ht]
\centering
\includegraphics[width=8.5cm]{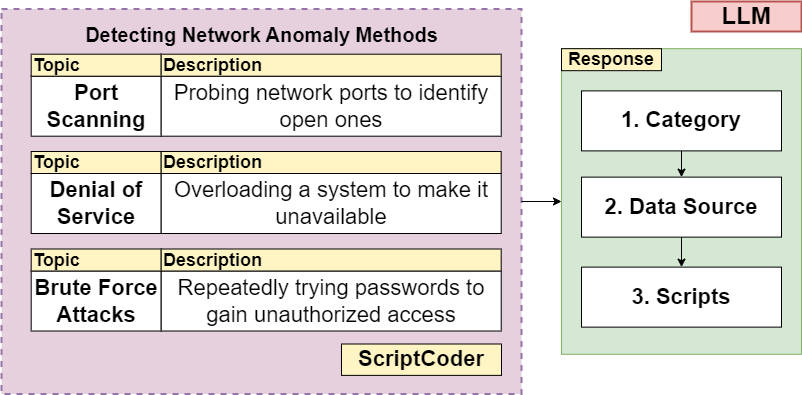}
\caption{A Skill-based Tool for Network Anomaly Detection}
\label{fig:NetworkAnomaly}
\end{figure}

\begin{figure}[ht]
\centering
\includegraphics[width=8.5cm]{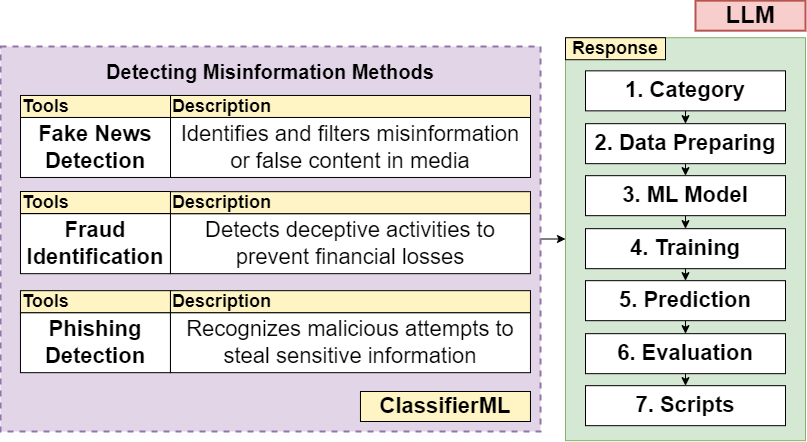}
\caption{A Skill-based Tool for Information Anomaly Detection}
\label{fig:InformationAnomaly}
\end{figure}

\subsection{Automated Scripting for Network Anomaly Detection}
Writing scripts using tools like Bash and AWK is essential for network anomaly detection. This is because these scripting tools enable automation of routine tasks, customization of data processing, and real-time analysis, which are crucial for efficiently managing network logs, detecting suspicious patterns, and responding to potential threats. However, students often struggle with scripting due to its complexity and the need to understand diverse network protocols and data sources. Our \textit{ScriptCoder} tool (Fig. \ref{fig:NetworkAnomaly}) assists students through a step-by-step workflow: it first categorizes the student's query and identifies key concepts such as specific network protocols or types of anomalies; then it determines the relevant data sources pertinent to the task, simplifying the process by focusing on essential components; finally, it generates the appropriate script based on the identified requirements, providing students with a ready-to-use tool and reinforcing their understanding of script construction. For example, if a student aims to detect port scanning activities within a network subnet using AWK, \textit{ScriptCoder} would classify the problem, specify relevant data sources like network logs, and generate the necessary AWK script. This structured approach ensures that students gain practical scripting experience while deepening their theoretical understanding.

\subsection{Machine Learning for Cybersecurity Threat Detection}
Machine learning plays a transformative role in cybersecurity by enabling the automatic detection and response to complex and evolving threats that traditional methods may fail to address. Writing code for machine learning models is therefore a critical skill, allowing students to analyze large datasets, identify malicious patterns, and adapt to new threats. Our \textit{MLClassifier} tool (Fig. \ref{fig:InformationAnomaly}) guides students through a comprehensive workflow for creating machine learning scripts in cybersecurity applications. It begins with problem categorization to ensure students understand the nature of the threat and the appropriate machine learning approach. The tool then assists in data loading and preprocessing tasks, such as cleaning and feature engineering. It helps students select an appropriate machine learning model based on the problem, provides instructions for training the model, and facilitates prediction generation on test data. \textit{MLClassifier} also outlines methods for evaluating model performance using metrics like accuracy, precision, and recall. Finally, it generates the complete Python script with detailed explanations for each step, serving both as a functional tool and an educational resource to help students understand each part of the script. For example, when tasked with detecting fake news, \textit{MLClassifier} would guide the student through categorizing the problem as a supervised learning task, selecting a dataset, pre-processing the data, choosing and training a SVM classification model, making predictions, and evaluating the results. The tool would then generate a full Python script that the student can use and modify, enhancing their coding skills and their understanding of machine learning applications in cybersecurity.

\section{Use Scenarios}\label{sec:use_scenarios}

To illustrate the practical applications and benefits of the developed system, we present a series of use cases that showcase its effectiveness in supporting both students and instructors. These scenarios demonstrate how the natural language assistant, powered by LLMs, can address the specific needs of various users, providing tailored educational support and resources in real-time.

\subsection{Use Case 1: Solving a Cryptography Problem for Student}

\begin{figure}[ht]
\centering
\includegraphics[width=8.5cm]{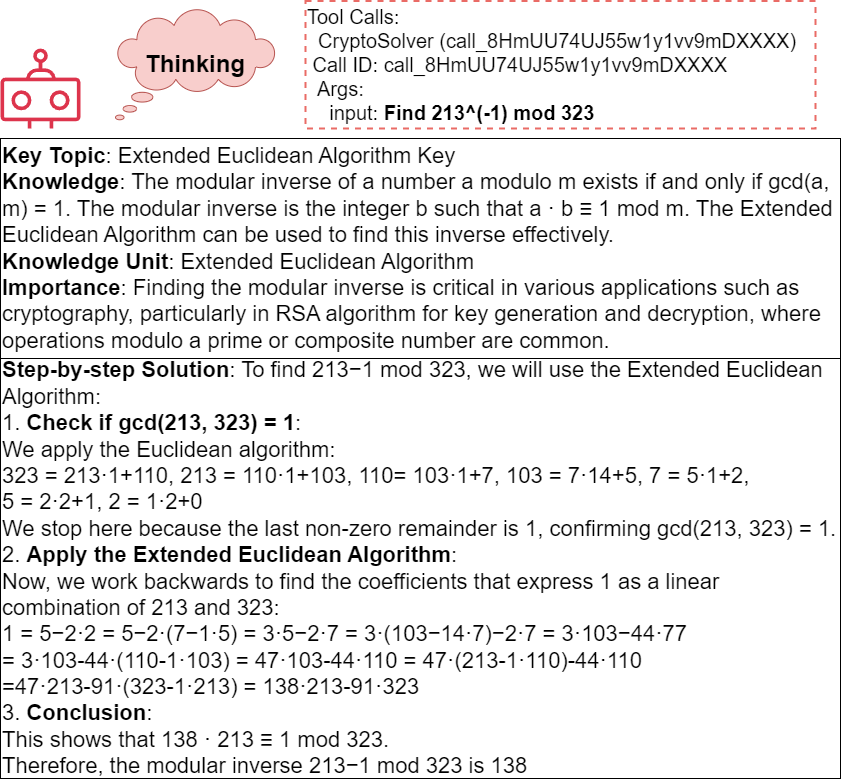}
\caption{Use Case 1: Solving a Cryptographic Problem.}
\label{fig:UseCase1}
\end{figure}

Maria is a part-time student and veteran. Balancing her academic responsibilities with a full-time job, Maria often finds it challenging to access educational resources during standard office hours. Her primary pain points include limited access to timely assistance and difficulties in grasping complex cryptographic concepts. As depicted in Fig. \ref{fig:UseCase1}, Maria engages with our AI-driven educational system by entering queries in natural language, such as ``Find 213\^{}(-1) mod 323". Upon receiving the query, the LLM-based agent quickly initiates a background process to select the most appropriate tool from its extensive knowledge base. During this phase, the user interface displays a "thinking" status, ensuring that the user is informed of the ongoing process without being burdened by the technical details. The agent then selects the specialized \texttt{CryptoSolver} tool, which is designed to handle cryptographic calculations, to compute the modular inverse problem. After generating the solution, the AI agent first verifies the accuracy and completeness of the response. It then presents the final answer to Maria in a two-step approach: 1) it provides a comprehensive overview of the relevant cryptographic principles, ensuring that Maria understands the foundational concepts. 2) it delivers a step-by-step solution to the query, guiding her through the problem-solving process in a clear and structured manner. This approach not only addresses Maria's immediate academic needs but also enhances her understanding of complex topics, thereby improving her overall learning experience.

\subsection{Use Case 2: Streamlining Course Preparation for Faculty}

Dr. Johnson is a professor tasked with developing a new course curriculum while managing a busy schedule filled with research and committee responsibilities. One of his main challenges is ensuring that his new course content aligns with the latest guidelines, such as the NICE Framework and CAE Knowledge Units (KUs), without requiring extensive time to manually curate these resources. Dr. Johnson accesses the AI-driven system to streamline his course preparation process. He begins by submitting a query through the natural language interface: ``Can you provide me the core knowledge units I should cover in Network Security class?" The system then initiates a background process where it scans its extensive knowledge base to identify relevant and up-to-date educational content. This process conducted entirely behind the scenes. Once the appropriate knowledge unit database is identified, the AI agent employs RAG techniques to extract the most relevant information. The system not only presents Dr. Johnson with the core knowledge units essential for his new course but also provides the source and URL for further exploration. This allows Dr. Johnson to validate the information and delve deeper into specific topics as needed, ensuring that his course remains both current and comprehensive.

\subsection{Use Case 3: Personalized Career and Course Guidance}

Alex, a cybersecurity student, seeks career opportunities that match his strengths in organization and attention to detail. However, navigating the vast array of potential roles and corresponding educational requirements can be overwhelming. His primary challenge lies in pinpointing the right career path and selecting appropriate courses that will prepare him for success in that path. Alex engages with the system through a series of conversational queries aimed at uncovering his ideal career path and the courses needed to achieve his goals. In the first round, Alex queries the system: ``What jobs in cybersecurity require strong organizational skills and attention to detail?" The system leverages the \texttt{kb\_career\_pathways} tool, which is tailored to provide career pathway guidance in line with the CAE Pathways guidelines. The tool generates a list of relevant cybersecurity roles, which the system then verifies for accuracy before presenting them to Alex. In the second round of interaction, satisfied with the initial response, Alex asks for course recommendations to align with his selected career path: ``Which courses should I take next?" The system, retaining the context from the previous conversational query, utilizes the \texttt{kb\_catalog\_advisor} tool to generate a list of courses focusing on virtual machines and organizational skills. This list is then verified and provided to Alex, complete with sources and URLs for further research, allowing him to explore the recommendations in more depth.

\section{Evaluation}


One of the known challenges of leveraging Generative LLMs is their tendency to suffer from hallucination, where the model generates plausible-sounding but factually incorrect or irrelevant information. This issue poses a risk to educational applications, where accuracy and reliability are paramount. Therefore, the evaluation of our proposed framework is a crucial aspect of this study, as it determines the effectiveness of utilizing Generative LLMs in cybersecurity education. To achieve a comprehensive assessment, we employed three key score metrics—helpfulness, correctness, and completeness. These metrics ensure that the model’s output not only provides relevant and valuable information but also aligns with accurate facts and logical reasoning. Helpfulness was assessed based on whether the answer effectively addressed the question, correctness was evaluated by checking the consistency of the response with a reference answer, and completeness measured whether all parts of the query were sufficiently answered.

Our evaluation utilized a dataset of 115 cybersecurity-related questions across three domains: knowledge-based, cryptography, and coding. These questions were sourced from industry interview questions, academic exercises, and practical problem-solving tasks. The diversity in question types ensures a comprehensive evaluation of the framework’s proficiency in both theoretical knowledge and technical skills. For detailed sources, please refer to our repository\footnote{https://github.com/tisage/CyberMentor/tree/main/eval}. To evaluate the performance, we utilized a LangChain prompt-based evaluation methodology\footnote{https://docs.smith.langchain.com/concepts/evaluation}, which involved an independent LLM to cross-validate the performance of the responses. 

\begin{table}[ht]
\centering
\caption{Framework Evaluation Across Different Source}
\label{tab:evaluation}
\begin{threeparttable}
\begin{tabular}{M{1.3cm} M{1.1cm} M{1.4cm} M{1.45cm} M{1.5cm}}
\hline
\textbf{Category} &
  \textbf{\# of Question} &
  \textbf{Helpfulness\tnote{*}} &
  \textbf{Correctness\tnote{*}} &
  \textbf{Completeness\tnote{*}} \\ \hline
Cybersecurity & 75      & 0.94  & 0.91  & 0.87  \\ \hline
Cryptography  & 10      & 0.87  & 0.63  & 0.89  \\ \hline
Coding        & 30      & 0.80  & 0.89  & 0.92  \\ \hline
\textbf{} &
  \textbf{115} &
  \textbf{0.85} &
  \textbf{0.83} &
  \textbf{0.90} \\ 
              & (Total) & (Avg) & (Avg) & (Avg) \\ \hline
\end{tabular}
\begin{tablenotes}
    \item[*] Each score is averaged over three independent experimental rounds.

\end{tablenotes}
\end{threeparttable}
\end{table}

The results (Table~\ref{tab:evaluation}) highlight that the framework achieved an overall average score of 0.85 for helpfulness, 0.83 for correctness, and 0.90 for completeness, signifying robust performance across different categories. The framework particularly excels in addressing cybersecurity-related interview questions, achieving a high correctness score (0.91) and an overall helpfulness score (0.94). These results validate the efficacy of the RAG methodology, which ensures that answers are sourced from extensive, relevant knowledge bases. By retrieving specific information tailored to the context of the question, the system successfully supports students in preparation for technical interviews, which are critical in their career development. In the coding-related tasks, the correctness score (0.89) suggests that it performs particularly well in practical, skill-based learning environments. This is especially notable for its role in guiding students through complex coding exercises and machine learning implementations, areas where skill proficiency is essential for both academic success and career readiness.

However, the framework exhibited some limitations in the cryptography domain, where it scored a correctness of 0.63. This lower performance is consistent with known difficulties LLMs face when tackling complex mathematical concepts. Current LLMs often struggle with tasks requiring precise mathematical reasoning, particularly in areas like cryptography. This gap indicates that while LLMs can be extremely effective in retrieving and presenting knowledge-based content, their capacity for handling intricate mathematical operations remains a challenge. As LLM architectures continue to evolve, improvements in handling such tasks are expected, potentially enhancing performance in mathematically intensive fields like cryptography. Despite these limitations, the overall completeness of the framework’s responses remains strong, with an average score (0.90) across all categories. The high completeness scores reflect the framework's ability to answer all components of a question comprehensively, ensuring students receive thorough explanations. This is particularly important in an educational setting, where fragmented or incomplete answers could lead to misunderstandings and hinder learning progress.

\section{Discussion and Conclusion}
\label{section:conclusion}
\subsection{Platform Adaptability and Extensibility}
\textit{CyberMentor}'s open-source and modular architecture facilitates extensive customization, enabling its adaptation to diverse learning contexts. This adaptability is primarily achieved through knowledge base expansion and tool extensibility. Our system supports seamless integration of external knowledge sources. Instructors and users can augment the system's knowledge base with various document formats, including PDFs, course materials, and research papers, thereby creating specialized learning environments tailored to specific curricula. For instance, an instructor teaching network security can incorporate relevant RFCs, cybersecurity standards documents, and research articles, ensuring contextually relevant responses to student queries. This dynamic knowledge base ensures access to up-to-date information and promotes curriculum alignment with current industry practices. Furthermore, the platform's modular design enables the integration of existing tools and the development of custom agent-based functionalities. Users can integrate established problem-solving tools or develop specialized agents to address specific learning objectives. For example, a user requiring network traffic analysis capabilities can integrate an existing network analysis tool or develop a custom agent leveraging specialized algorithms. This extensibility empowers users to create highly customized learning environments that cater to diverse learning styles and specific skill development needs. Moreover, this modularity fosters community-driven development by encouraging contributions of new tools and agents, benefiting the broader educational community. This open and extensible architecture makes \textit{CyberMentor} readily adaptable to various learning systems beyond cybersecurity, extending its applicability to other STEM disciplines and any field requiring knowledge retrieval and problem-solving support.

\subsection{Discussion}
The integration of AI-driven systems into cybersecurity education presents a significant opportunity to provide scalable and personalized support to all students, particularly those in non-traditional settings. However, AI technologies, including generative models like LLMs, still exhibit limitations, especially in tasks that require precise mathematical reasoning and logical coherence. While RAG enhances information retrieval, techniques such as chain-of-thought (CoT) reasoning offer potential to bridge gaps in AI's ability to handle complex reasoning tasks\cite{li2024chainthoughtempowerstransformers}. Furthermore, other research highlights the importance of scaling inference-time computation in LLMs, which could be leveraged in educational contexts to dynamically adjust AI assistance based on task complexity\cite{snell2024scaling}. 
However, despite these advances in AI technologies, the role of human mentorship remains indispensable. The behavioral science model emphasizes the importance of human interaction in achieving organizational goals, which can be analogously applied to the educational context\cite{block2014mystery}. Therefore, a hybrid approach that integrates AI-driven personalized learning with human mentorship is essential for creating an inclusive and adaptive educational ecosystem.

\subsection{Conclusion}
This paper has introduced \textit{CyberMentor}, an AI-driven framework designed to meet the diverse needs of students in cybersecurity education. The educational contributions of our work are threefold. First, it bridges the gap between the evolving demands of cybersecurity education and the limited availability of educational resources. Second, by integrating a dual-component approach consisting of a knowledge base and a skill base, enhanced by advanced LLM agents, the framework addresses critical pain points in cybersecurity education. Specifically, it provides access to up-to-date information, a balanced curriculum, career-specific learning support, and enhanced practical skill development through scalable, extensible, personalized 24/7 learning system. Third, our system is open-source and fully customizable, streamlining course preparation for faculty and ensuring alignment with industry standards, thus improving the overall quality and relevance of course offerings.

Currently, our framework offers fully functional demonstration capabilities but with a limited knowledge base and tools for cybersecurity courses. Future work will focus on expanding the knowledge base with graph database, refining AI capabilities, and incorporating advanced techniques such as CoT for mathematical reasoning. Pilot studies and feedback from users will be crucial for further development, ensuring that this framework continues to meet the evolving needs of cybersecurity education. We plan to conduct pilot studies and collect school-wide feedback from students and faculty to iteratively refine the system. Our goal is to ensure that \textit{CyberMentor} evolves into a valuable resource, supporting not only cybersecurity education but also other STEM disciplines within the increasingly complex landscape of higher education.

\bibliographystyle{IEEEtran}
\bibliography{references_cyber}

\newpage



\end{document}